\documentclass[aps,pra,twocolumn,longbibliography,superscriptaddress]{revtex4-2}
\usepackage{amsmath}
\usepackage{amssymb}
\usepackage{mathtools}
\usepackage{graphicx}
\usepackage{comment}
\usepackage{tikz}
\usepackage[normalem]{ulem}
\usepackage{bm}
\usepackage{isotope}
\usepackage{lineno}
\usepackage[T1]{fontenc}
\usepackage[utf8]{inputenc}
\usepackage[titletoc, title]{appendix}

\usepackage{hyperref}
\hypersetup{
   pdfpagemode=None, 
   pdfstartpage=1,
   pdfmenubar=true,
   pdftoolbar=true,
   colorlinks = true,
   linkcolor=blue,
   citecolor=blue,
   urlcolor=blue,
   bookmarksopen=false
 }

\usepackage[normalem]{ulem}

\newcommand{\kF}{k_F}
\newcommand{\eF}{\varepsilon_F}

\definecolor{darkblue}{rgb}{0.,0.24,0.51}
\definecolor{britishracinggreen}{rgb}{0.0, 0.26, 0.15}
\definecolor{darkgreen}{rgb}{0,0.60,.2}

\newcommand{\abs}[1]{\lvert{#1}\rvert}            

\newcommand{\prlsection}[1]{\textit{#1} ---}            

\begin{document}
\title{Dynamical signature of vortex mass in Fermi superfluids}

\author{Andrea Richaud}
\affiliation{Departament de F\'isica, Universitat Polit\`ecnica de Catalunya, Campus Nord B4-B5, E-08034 Barcelona, Spain}
\author{Matteo Caldara}
\affiliation{Scuola Internazionale Superiore di Studi Avanzati (SISSA), Via Bonomea 265, I-34136, Trieste, Italy}

\author{Massimo Capone}
\affiliation{Scuola Internazionale Superiore di Studi Avanzati (SISSA), Via Bonomea 265, I-34136, Trieste, Italy}
\affiliation{CNR-IOM Democritos, Via Bonomea 265, I-34136 Trieste, Italy}

\author{Pietro Massignan}
\affiliation{Departament de F\'isica, Universitat Polit\`ecnica de Catalunya, Campus Nord B4-B5, E-08034 Barcelona, Spain}

\author{Gabriel Wlaz\l{}owski}
\affiliation{Faculty of Physics, Warsaw University of Technology, Ulica Koszykowa 75, 00-662 Warsaw, Poland}
\affiliation{Department of Physics, University of Washington, Seattle, Washington 98195--1560, USA}

\date{\today}
	
\begin{abstract}
Quantum vortices are commonly described as funnel-like objects around which the superfluid swirls, and their motion is typically modeled in terms of massless particles. Here we show that in Fermi superfluids the normal component confined in the vortex core provides the vortex with a finite inertial mass. This inertia imparts an unambiguous signature to the dynamic behavior of vortices, specifically manifesting as small-amplitude transverse oscillations which remarkably follow the prediction of a simple point-like model supplemented by an effective mass. We demonstrate this phenomenon through large-scale time-dependent simulations of Fermi superfluids across a wide range of interaction parameters, at both zero and finite temperatures, and for various initial conditions. Our findings pave the way for the exploration of inertial effects in superfluid vortex dynamics.

\end{abstract}

\maketitle

\prlsection{Introduction} The observation of tiny effects may reveal noteworthy phenomena and even overturn well-established physical theories, a spectacular example being the discovery of the mass of neutrinos.
Around 25 years ago, in fact, high-precision experiments revealed the so-called neutrino flavour oscillations, which proved that neutrinos have masses that are certainly tiny, but definitely nonzero, contrary to the predictions by the Standard Model of particle physics. 
Not only did this discovery deserve the Nobel Prize in Physics in 2015~\cite{Nobel2015}, but it also called for a huge conceptual leap which led to the search for physics beyond the Standard Model. 

\begin{figure}[b!]
    \centering
    \includegraphics[width=0.9\columnwidth]{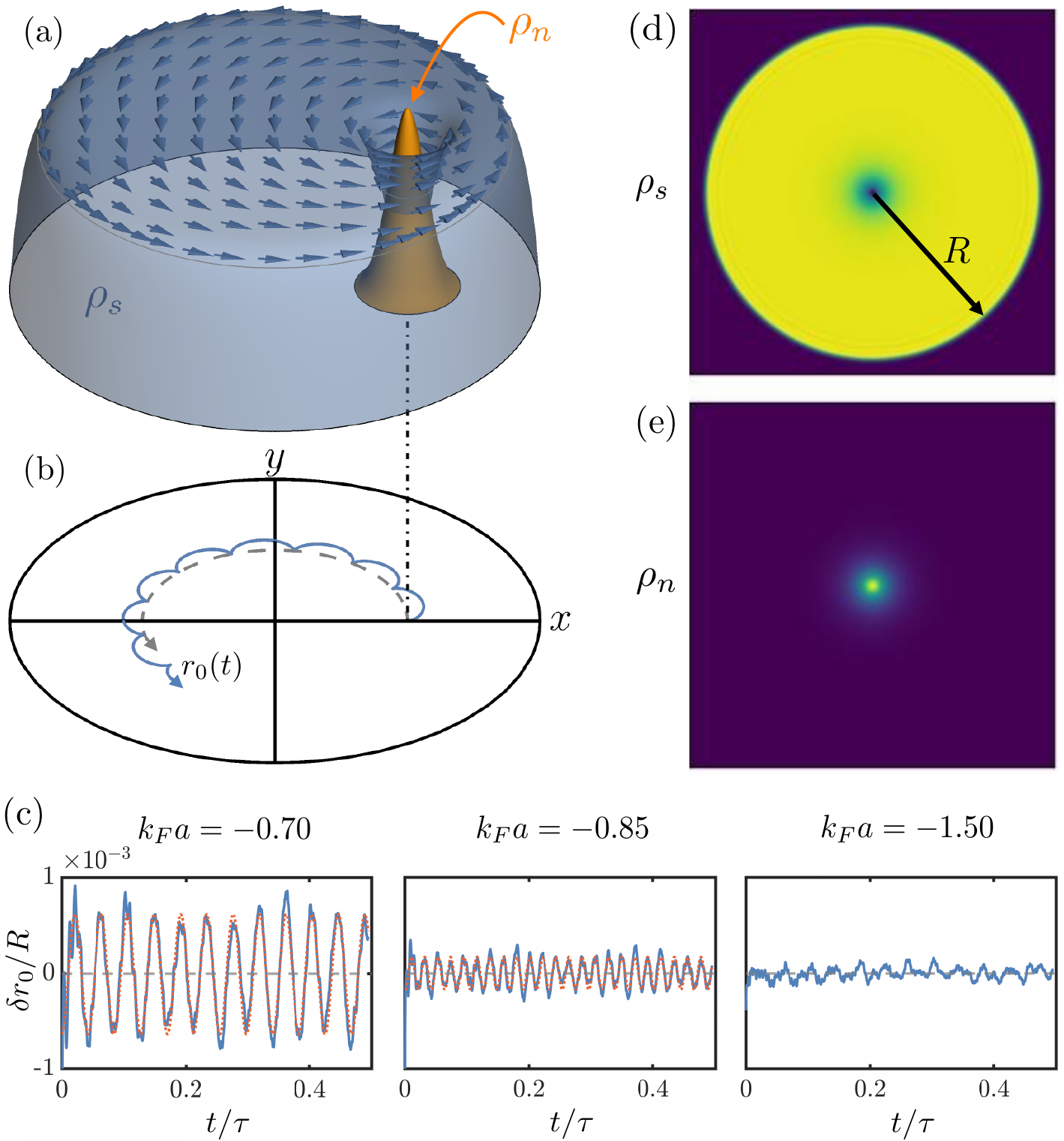}
    \caption{(a) Pictorial illustration of a superfluid (blue) confined in a circular trap and hosting an off-centered vortex whose core is filled by some normal component (orange). Panel (b) shows the trajectory (solid blue line) of the massive vortex characterized by small-amplitude radial oscillations around the circular trajectory of a massless vortex (dashed line). (c) Evolution of the radial coordinate of a vortex orbiting within a Fermi superfluid at zero temperature (blue solid line, the red dashed line is a sinusoidal fit) as the parameter $\kF a$ is changed. Our simulations start with a static vortex at  $r_0(t=0)=0.54\, R$. Moving from the deep BCS regime to the strongly-attractive regime (i.e., from the left to the right panel), the oscillation frequency increases, while the amplitude is reduced. 
    The superfluid (d) and normal (e) densities as extracted from static calculations of a vortex in a container of radius $R=90\, k_F^{-1}$ at $\kF a=-0.7$. Yellow (blue) color corresponds to high (zero) density. 
    }
\label{fig:Zero-temperature-results}
\end{figure}

In this Letter we explore a similar scenario for quantum vortices, which are fundamental excitations found across various quantum fluids. The simplest, yet very popular and useful, theoretical model describes vortices as massless funnel-shaped objects, devoid of any inertial effects~\cite{Kim2004,Hernandez2024}. 
Recent theoretical works highlighted that a non-zero mass has a profound dynamical signature, namely rapid transverse oscillations of the vortex~\cite{Richaud2021,Dambroise2024} which have no counterpart in the massless case. As an example, a massless vortex confined in a circular bucket travels in an ideal circular orbit, while a massive vortex exhibits additional radial oscillations, as shown schematically in Figs.~\ref{fig:Zero-temperature-results}(a-b).

The very existence of such vortex mass and its origin have been foreshadowed in several works scattered through the decades (see Refs.~\cite{Baym1983,Simula2018} and references therein).  
These pioneering studies were unavoidably limited to phenomenological approaches that required different a priori assumptions and that led to several distinct estimates of the vortex effective mass. 
In fermionic superfluids including superconductors, the appearance of a mass has been tentatively ascribed either to the normal component trapped in the core~\cite{Kopnin1991,Kopnin1998,Simanek1995,Duan1993,Volovik1998,Kopnin2002,Simula2018,Suhl1965,Simanek1994,Sonin2013}, or to the bound fermionic states in the core identified by Caroli, De Gennes, and Matricon~\cite{Caroli1964}.
Yet, different mechanisms with conflicting predictions have been proposed, and consensus is lacking on both the microscopic origin of the mass and, most importantly, on its impact on the observable properties of the vortex. 

In this work, we fill this gap through large-scale numerical calculations which represent the current gold standard for modeling the quantum dynamics of Fermi superfluids~\cite{Pecak2024}. 
Our analysis is based on microscopic calculations for a two-component fermionic superfluid. The most suitable tool to study the dynamics of a vortex in a microscopic framework is the time-dependent Hartree-Fock-Bogoliubov theory, which for fermionic superfluids is described by the time-dependent Bogoliubov-de Gennes (BdG) equations. This method was originally used to study the static properties of single vortices in the crossover from Bardeen-Cooper-Schrieffer (BCS) type superfluidity to a Bose-Einstein condensate (BEC) of tightly bound pairs~\cite{Nygaard2003,Chien2006,Sensarma2006,Simonucci2013,Simonucci2019}.
Over the last decades, it benefited from the development of the time-dependent Superfluid Local Density Approximation (SLDA), an {\it ab initio} method based on Density Functional Theory (DFT)~\cite{Bulgac2007,Bulgac2012}. The swift progress of DFT for superfluids is strongly tied to the spectacular advances in computational physics, and more specifically in High-Performance Computing, which today allow to investigate superfluid dynamics across the BCS-BEC crossover at unprecedented scale~\cite{Wlazlowski:10.1093}. 
The successful application of SLDA to the study of quantum vortices in Fermi gases~\cite{Bulgac2011,PhysRevLett.91.190404,PhysRevLett.112.025301,Magierski2022,Wlaszlowski_sound} has led to remarkable agreement with experiments related to vortex nucleation process~\cite{PhysRevLett.120.253002}, lattice formation~\cite{PhysRevA.104.053322} or vortex dipole collisions~\cite{Barresi2023}. By combining this powerful numerical technique with advanced many-body analyses and an effective point-vortex model, we unambiguously detect and characterize small transverse oscillations in the vortex trajectory, the hallmark of its inertial mass, which we attribute to the normal component trapped in the core. 

\prlsection{Massless and massive quantum vortices}
We consider an effectively two-dimensional (2D) uniform superfluid trapped in a disk of radius $R$ and hosting a point-like and massless vortex at position $\bm{r}_0$. The vortex dynamics is essentially determined by the boundary, which is modeled by an oppositely-charged image vortex located at position $\bm{r}_0^\prime=(R/r_0)^2 \, \bm{r}_0$ and ensuring no-flow boundary conditions. In the absence of dissipation, the considered vortex is propelled by its image and moves with the local superfluid velocity $\bm{v}_s$ according to the first-order equation of motion 
\begin{equation}
    \dot{\bm{r}}_0 \equiv \bm{v}_s= \hat{z}\times \frac{\hbar}{m_b} \frac{\bm{r}_0}{R^2-r_0^2},
    \label{eq:First_order_ODE}
\end{equation}
where $m_b$ is the mass of the (effective) boson constituting the superfluid and $\hat{z}$ is the unit vector parallel to the vortex axis~\cite{Donnelly1991}. In this framework, a quantum vortex can only travel along circular orbits with radius $r_0$, shown by a gray dashed line in Fig.~\ref{fig:Zero-temperature-results}(b).   

The scenario changes quite dramatically if the vortex has a non-zero inertial mass $M_c$ \cite{Baym1983,Turner2009,Richaud2021}. In this case, in fact, the vortex no longer moves with the local superfluid velocity, but follows the Newton-like equation of motion
\begin{equation}
    M_c \ddot{\bm{r}}_0 = 2\pi\hbar\bar{\rho}_s\left(\bm{v}_s -\dot{\bm{r}}_0 \right)\times \hat{z}\label{eqn:MPVM}
\end{equation}
regardless of the microscopic origin of $M_c$ ($\bar{\rho}_s$ represents the 2D bulk superfluid density). In stark contrast to the massless case, a massive vortex exhibits small-amplitude transverse oscillations superimposed to the previously described uniform circular orbit~\cite{Richaud2021}, which are depicted by the solid blue line in Fig.~\ref{fig:Zero-temperature-results}(b). The frequency
\begin{equation}
    \omega = \frac{1}{\tau}\frac{2}{\mathfrak{m}}\sqrt{1-\mathfrak{m} \frac{2-\tilde{r}_0^2}{\left(1-\tilde{r}_0^2\right)^2}}
    \label{eq:omega}
\end{equation}
where $\tau= m_bR^2/\hbar $ is a convenient unit of time, $\tilde{r}_0=r_0/R$ is the dimensionless radial position, and $\mathfrak{m} = M_c/M_s$ represents the core-superfluid mass ratio ($M_s=m \int \rho_s(\bm{r})\,\mathrm{d}\bm{r}$ denotes the total mass of the superfluid). In the limit $R \to \infty$, the oscillation frequency remains well defined and approaches $\omega = 2\pi\hbar\bar{\rho}_s / M_c$, confirming that the vortex mass $M_c$ retains its physical relevance in the hydrodynamic limit. Moreover, in the small-mass limit such frequency displays the same scaling $M_c^{-1}$ as the cyclotron frequency~\cite{Munoz_de_las_Heras2020}. This is due to the Magnus-like form of the right-hand side of Eq.~(\ref{eqn:MPVM}) and is consistent with the formal analogy between quantum vortices and electric charges~\cite{Caldara2023}.
In our analysis we measure the oscillation frequency $\omega$ and then we extract the vortex mass $\mathfrak{m}$ through Eq.~(\ref{eq:omega}). This approach is free of any assumptions regarding the source of the vortex inertia. The mass $M_c$ is an intrinsic property of the vortex which does not depend on the system details. The finite-size system we consider is a convenient and experimentally realistic platform to unveil the vortex inertial effect with unprecedented detail. 

\prlsection{Weighting a vortex from its trajectory} 
Here we present the results of large-scale numerical simulations of the motion of a single off-centered quantum vortex in a homonuclear two-component fermionic superfluid (with atomic mass $m$) confined in a cylindrical container. We first prove that vortex trajectories present small radial oscillations compatible with Eq.~(\ref{eqn:MPVM}). Then, relying on both static and time-dependent DFT approaches (see Supplemental Material~\cite{SM} for details), we demonstrate that the mass of the normal component trapped in the core of quantum vortices is compatible with the value inferred from the transverse-oscillation frequency (\ref{eq:omega}). Note that both the superfluid and normal components arise naturally within the DFT framework without any \emph{a priori} assumption.  
We characterize the interaction strength by the product of the three-dimensional $s$-wave scattering length $a$ and the Fermi wave-vector $k_F=(3\pi^2\bar{\rho})^{1/3}$, with $\bar{\rho}$ the 3D bulk density.
We focus on the BCS regime $(k_Fa)^{-1} \lesssim -1$, where a reliable energy density functional exists~\cite{Boulet2022} and the vortex mass is sufficiently large to be measured.
 Additionally, we comment on the strongly interacting regime $-1 \lesssim (k_Fa)^{-1} \leq 0$, where the vortex mass drops below detectable values, and ultimately approaches zero in the deep BEC regime. 

\begin{figure}[t!]
    \centering
    \includegraphics[width=0.9\columnwidth]{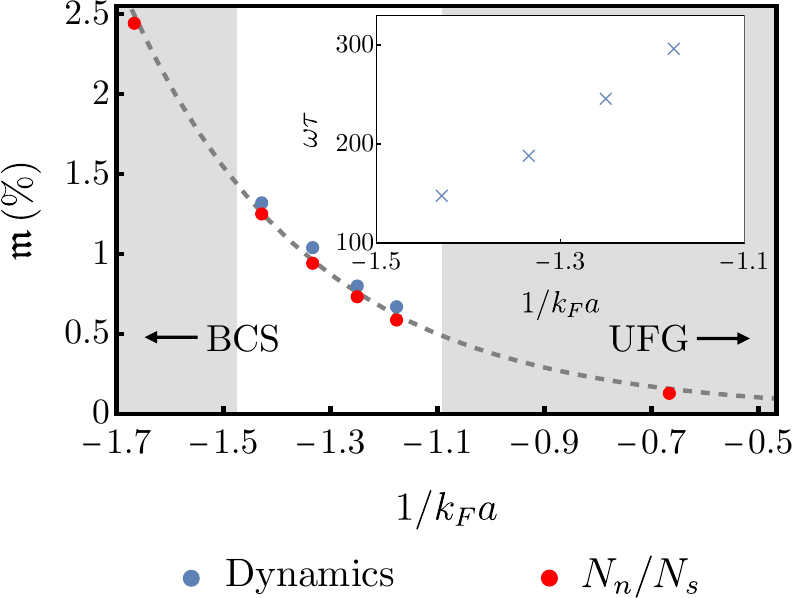}
    \caption{Dependence of the vortex mass $\mathfrak{m}$ on the interaction strength as obtained from the frequency $\omega$ of transverse oscillations through Eq.~(\ref{eq:omega}) (blue dots) and from the ratio $N_n/N_s$ (red dots). The gray dashed line corresponds to Eq.~(\ref{eq:m_equal_const_xi_squared}). Inset: $\omega(k_Fa)$ extracted from time-dependent DFT simulations. Gray areas correspond to values of $k_Fa$ for which $\omega$ cannot be extracted from the trajectory in a reliable way. 
    }
    \label{fig:Fig3}
\end{figure}

Fig.~\ref{fig:Zero-temperature-results}(c) shows the time-evolution of the vortex radial position, $r_0(t)=\langle r_0 \rangle + \delta r_0(t)$, for three different values of the interaction parameter $k_Fa$, obtained from time-dependent DFT simulations at zero-temperature ($\langle r_0 \rangle$ is the average radial position, and $\delta r_0(t)$ is the fluctuation around it). The first two panels, corresponding to $k_Fa=-0.70$ and $k_Fa=-0.85$, feature clear and regular oscillations (solid blue lines) that represent the hallmark of the non-zero inertial mass of the simulated vortex.
These oscillations, despite their very small amplitude $\sim 0.1 dx$ (with $dx$ the mesh-element length), can be detected through a vortex tracking algorithm with sublattice resolution~\cite{SM}. We also thoroughly checked that varying the initial velocities of the massive vortex results in oscillations of different amplitudes but unchanged frequency, as it is expected in the framework of linear response for inertial particles~\cite{SM}. Also, with various tests, we ruled out the possibility of numerical artifacts in the analyzed signal $r_0(t)$~\cite{SM}. Importantly, the oscillations are undamped within our observation window, indicating the absence of significant energy transfer from the vortex motion to other excitation channels such as phonon or surface modes, which would otherwise lead to measurable dissipation. This suppression is consistent with two factors: \textit{i)} the phonon wavelength $\lambda = 2\pi c_0/\omega$ exceeds the system size ($\lambda/R > 2.2$), inhibiting efficient emission~\cite{Fetter2001} of sound waves ($c_0$ is the sound velocity~\cite{KetterleZwierleinReviewFermi}), and \textit{ii)} the hard-wall confinement that we employ implies that no surface modes can exist, in contrast to harmonically-trapped systems where surface modes are naturally present and easily couple to Kelvin modes (see, e.g. Ref.~\cite{Bretin2003}). A fit of the regular oscillations with sinusoidal functions (red dashed lines) gave us the relevant frequency $\omega$. The latter turned out to steadily increase upon moving from the deep BCS ($1/k_Fa\ll-1$) towards the unitary Fermi gas (UFG) regime ($1/k_Fa=0$), see the inset of Fig.~\ref{fig:Fig3}. 

In parallel, we computed the superfluid $\rho_s(\bm{r})$ and the normal $\rho_n(\bm{r})$ density distributions of the system. To achieve this, we centered the vortex at the origin and performed static DFT calculations using the same microscopic parameters as in the time-dependent simulations. The calculations produce distributions of the total number density $\rho(\bm{r})$, current density $\bm{j}(\bm{r})$, and (complex) superfluid order parameter $\Delta(\bm{r})$. The current can be decomposed as $\bm{j}=\rho_s \bm{v}_s + \rho_n \bm{v}_n$, where $\bm{v}_s$ and $\bm{v}_n$ are the velocities of the superfluid and normal component, respectively (the explicit dependence on $\bm{r}$ has been omitted). The superfluid velocity can be expressed as $\bm{v}_s=\frac{\hbar}{m_b}\nabla\varphi$, being $\varphi$ the phase field associated to the order parameter $\Delta=|\Delta|e^{i\varphi}$ and $m_b=2m$ the mass of the Cooper pair resulting from quantum correlations between two fermions with opposite spins.
We measure the current response $\bm{j}$ to a phase-imprint $\varphi$ with a $2\pi$-winding around the origin. Since the phase imprint only induces superflow ($\bm{v}_n=\bm{0}$), the superfluid and normal densities are extracted from $\bm{j}=\rho_s \frac{\hbar}{m_b} \bm{\nabla} \varphi$ and $\rho_n = \rho - \rho_s$, respectively.
The panels (d) and (e) of Fig.~\ref{fig:Zero-temperature-results} show the 2D distributions for $\rho_s$ and $\rho_n$, respectively. A clear density suppression in $\rho_s$ represents the core of the superfluid vortex, while the corresponding peak in $\rho_n$ indicates that it is filled with the normal component. 
We quantify this observation by computing the number of particles that effectively belong to either the superfluid  
    $N_s = \int \rho_s(\bm{r})\, \mathrm{d}\bm{r}$
or the normal 
    $N_n = \int \rho_n(\bm{r})\, \mathrm{d}\bm{r}$
components (their sum equals the total number of atoms $N\approx 1.3\times 10^4$). As illustrated by the red dots in Fig.~\ref{fig:Fig3}, the ratio $N_n/N_s$ monotonically decreases as a function of $1/k_Fa$, indicating that the vortex core is more filled in the BCS regime compared to the strongly-interacting regime. The blue dots in the same figure show the value of $\mathfrak{m}$ inferred from the time-dependent simulations by extracting the oscillation frequency $\omega$ of the vortex motion and inverting Eq.~(\ref{eq:omega}). We note that the uncertainty in the fitted frequency is very small, with a relative error $\Delta\omega/\omega \approx 4 \times 10^{-4}$, which translates into a relative uncertainty $\Delta\mathfrak{m}/\mathfrak{m}$ in the vortex mass of less than $5 \times 10^{-4}$. The corresponding error bars would therefore be smaller than the symbol size and are not shown.
Such dynamic protocol suffers from some intrinsic limitations. Moving towards unitarity (right gray box in Fig.~\ref{fig:Fig3}) the vortex becomes increasingly lighter, and both the period and the amplitude of the radial oscillations become too small to be resolved [see right panel in Fig.~\ref{fig:Zero-temperature-results}(c)]. Conversely, in the deep BCS regime (left gray box) the vortex core becomes increasingly large, to the point where the massive point-vortex model \cite{Richaud2021} is no longer applicable.
The excellent quantitative agreement between the static and dynamic indicators confirms that the normal component trapped inside the core region provides the vortex with a finite inertial mass. This is our main result. While we are not able, in principle, to disentangle a hydrodynamic contribution~\cite{Baym1983} to the vortex mass, the close agreement with the localized normal component suggests that such a contribution would be, at least, subleading in the BCS regime.

Both microscopic and phenomenological approaches~\cite{Kopnin1991,Duan1993,Volovik1998,Kopnin2002,Simula2018} predict that the vortex mass is proportional to the area of the core ($\propto \xi^2$, where $\xi$ is the coherence or healing length). 
However there is no agreement on the prefactor, which strongly depends on the specific model used for the vortex core. To test the validity of this statement, we plot in Fig.~\ref{fig:Fig3} the relation
\begin{equation} \label{eq:m_equal_const_xi_squared}
    \mathfrak{m}= \alpha \times (\xi/R)^2,
\end{equation}
(see gray dashed line) with $\alpha$ a suitable fit constant and 
    $\xi= \frac{\hbar^2 k_F}{m \pi \bar{\Delta}}$
the BCS coherence length, being $\bar{\Delta}$ the bulk value of the order parameter $\Delta$. 
The agreement of Eq.~(\ref{eq:m_equal_const_xi_squared}) with our numerical results extracted from both the cyclotron dynamics and the ratio $N_n/N_s$ is remarkable. The extracted fit parameter is $\alpha= 1.5(1)$, hence of the order of unity. It is also worth mentioning that our DFT numerics shows that there is no direct relation between the normal density $\rho_n$ and the Caroli-de Gennes-Matricon states bound to the vortex core (see~\cite{SM} for a detailed discussion). 

\smallskip
\noindent

\begin{figure}[t!]
    \centering
    \includegraphics[width=0.9\columnwidth]{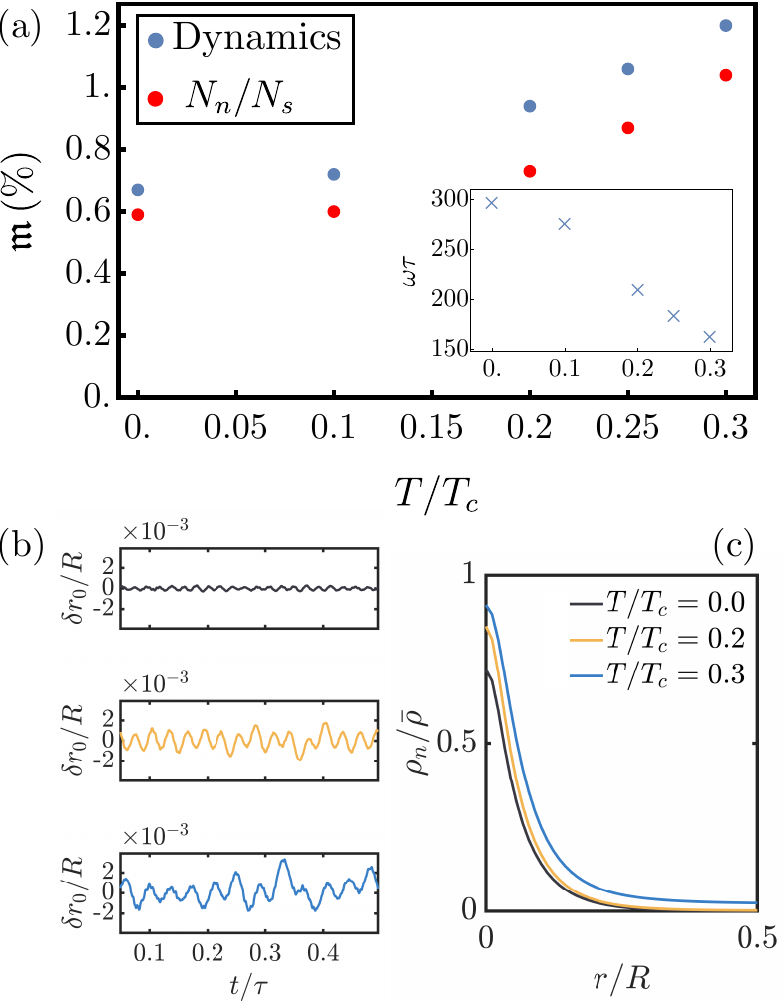}
    \caption{Dependence of the vortex mass $\mathfrak{m}$ on temperature $T$ for fixed $k_Fa=-0.85$ and fixed initial conditions. (a) Comparison from the values inferred from the oscillation frequency  $\omega$ (blue dots) and the ratio $N_n/N_s$ (red dots). Error bars would be smaller than the symbols. Inset: $\omega(T)$ extracted from the time-dependent DFT simulations. (b) Time evolution of the vortex radial position: oscillations are less frequent and wider at higher temperatures [lines are color-coded as in panel (c)]. (c) Temperature initially increases the normal component in the vortex core, and, for larger values, it also enhances $\rho_n$ in the bulk.}
    \label{fig:Finite-T-results}
\end{figure}

\prlsection{Temperature-enhanced vortex mass}
DFT is a versatile method that allows for the inclusion of finite temperature effects. We now study the role of temperature, which is a crucial step to validate the robustness of the discussed phenomenology and its observability in on-going experiments. Thermal fluctuations are expected to induce extra normal component, first in the vortex core where the energy gap $\Delta$ is suppressed, and next in the bulk. The amount of thermally-induced normal component localized within the vortex core should then lead to an enhanced vortex inertial mass and, consequently, slower transverse oscillations. Indeed, this is what we observe, as shown in Fig.~\ref{fig:Finite-T-results} for a fixed interaction parameter $k_F a = -0.85$. As before, we present the comparison of the vortex mass extracted either from the ratio $N_n/N_s$ or from the transverse-oscillation frequency $\omega$ by inverting Eq.~(\ref{eq:omega}). The temperature dependence of both these indicators is shown in Fig.~\ref{fig:Finite-T-results}(a) [as a technical note, to compute $N_n$ we removed, if present, the bulk value of $\rho_n$]. Moreover, Fig.~\ref{fig:Finite-T-results}(b) compares the vortex radial oscillations at zero and finite temperature. In the latter case, the oscillations are wider and slower, consistent with the larger amount of normal component localized in the vortex core, as shown in Fig.~\ref{fig:Finite-T-results}(c). Notice, in particular, that the revealed increasing trend of the oscillation amplitude with temperature for $T \lesssim 0.3\,T_c$ ($T_c$ being the critical temperature of the superfluid-to-normal phase transition) seems promising towards the possibility of observing such oscillatory vortex dynamics in state-of the art experiments~\cite{Kwon2021,Hernandez2024,Cabrera2024}.  
The oscillation becomes irregular once a significant amount of the normal component is induced in the bulk, thus preventing the extraction of the cyclotron frequency $\omega$. The fact that we observe undamped oscillations for $T\lesssim 0.3\,T_c$ indicates that dissipation can be safely neglected in these conditions, in full agreement with recent measurements~\cite{Grani2025}. At higher temperatures, the simple model~(\ref{eqn:MPVM}) should instead be supplemented with extra terms accounting for dissipative forces.

\prlsection{Conclusions and Outlook}
We have shown that vortices in Fermi superfluids are naturally massive because of the presence of the normal component localized in the core. The fraction of normal component has been found to increase as the interparticle interactions decrease, determining an increase of the period of the transverse oscillations, which thus constitutes a reliable observable to quantify the effective vortex mass. The above scenario emerges on the one hand from large-scale numerical simulations of many-body Fermi systems, and, on the other hand, it is well described by a simple yet powerful generalized point-vortex model that accounts for the intrinsic inertia of the vortex. Furthermore, we demonstrated that finite-temperature effects enhance the visibility of the vortex inertial signature.

The ability to characterize the intrinsic mass of a vortex opens new paths to explore the intriguing dynamical and statistical behavior of vortices in Fermi superfluids. The single-vortex system studied here already constitutes the ideal platform for addressing critical open questions such as the origin of dissipation in quantum vortex motion~\cite{Hall1956,Kwon2021,Barresi2023,Grani2025}, vortex-sound interactions~\cite{Lundh2000,Parker2004,Kwon2021}, and the microscopic origin of the normal component. Our findings are ideally suited to be explored in ongoing precision experiments investigating vortex dynamics in uniform superfluids across the BCS-BEC crossover \cite{Kwon2021,Hernandez2024,Cabrera2024,Grani2025} in both spin-balanced and spin-imbalanced Fermi gases~\cite{Magierski2022}. Our results have immediate implications for 2D superfluids, where vortex inertia is expected to impact a range of phenomena, including Tkachenko modes in vortex lattices, shifts in the Berezinskii-Kosterlitz-Thouless transition temperature, and the growth rate of the superfluid Kelvin-Helmholtz instability~\cite{Baggaley2018,Giacomelli2023,Hernandez2024,Caldara2024}. In multi-component systems or those with impurities trapped in vortex cores, the effective vortex mass is further enhanced. Finally, the vortex mass is also expected to affect the dynamics of three-dimensional vortex lines, particularly the behavior of Kelvin waves at scales comparable to the vortex core size~\cite{Barenghi2006,Simula2018}, which play a key role in the energy cascade of quantum turbulence~\cite{Barenghi2023,Reeves2022,Wlazlowski:10.1093}.

\smallskip
\prlsection{Acknowledgments}
We thank A. L. Fetter for a critical reading of the manuscript and useful suggestions and A. Recati for stimulating discussions.
A.R. received funding from the European Union’s Horizon research and innovation programme under the Marie Skłodowska-Curie grant agreement \textit{Vortexons} (No.~101062887), and from the ``La Caixa'' Foundation through the Junior Leader Fellowship LCF/BQ/PR25/12110014. A.R. and P.M. acknowledge financial support by the Spanish Ministerio de Ciencia, Innovación y Universidades (grant PID2023-147469NB-C21, financed by MICIU/AEI/10.13039/501100011033 and FEDER-EU).
P.M. further acknowledges support by the {\it ICREA Academia} program. 
M. Cal., A. R. and P. M. would like to thank the Institut Henri Poincaré (UAR 839 CNRS-Sorbonne Université) and the LabEx CARMIN (ANR-10-LABX-59-01) for their support. G.W. acknowledges the National Science Centre, Poland, for support under grant UMO-2022/45/B/ST2/00358. 
M. Cap. acknowledges financial support of MUR via PRIN 2020 (Prot. 2020JLZ52N 002) programs, and by the European Union - NextGenerationEU through PRIN 2022 (Prot. 20228YCYY7), National Recovery and Resilience Plan (NRRP) MUR Project No. PE0000023-NQSTI, and No. CN00000013-ICSC.
We acknowledge the Polish high-performance computing infrastructure PLGrid for awarding this project access to the LUMI supercomputer, owned by the EuroHPC Joint Undertaking, hosted by CSC (Finland) and the LUMI consortium through PLL/2023/04/016476.

\smallskip
\prlsection{Author contributions} A.R., M.Cal. and G.W. conceived the study. G.W. prepared and performed the numerical simulations. A.R., M.Cal, P.M. and G.W. analyzed the output of numerical simulations. All authors contributed to the interpretation of the results and to the writing of the manuscript.

\smallskip
\prlsection{Data availability} Data underlying the study and detailed instructions on how to reproduce the results are available via the Zenodo repository~\cite{Zenodo}.

%


\vspace{1cm}
\begin{center}
\textbf{
Supplementary Information}\\
\end{center}

\setcounter{equation}{0}
\setcounter{figure}{0}
\setcounter{table}{0}
\setcounter{section}{0}
\setcounter{page}{1}
\makeatletter
\renewcommand{\theequation}{S.\arabic{equation}}
\renewcommand{\thefigure}{S\arabic{figure}}
\renewcommand{\thetable}{S\arabic{table}}
\renewcommand{\thesection}{S.\arabic{section}}

\section{Methods}

\subsection{DFT simulations}
\label{sub:DFT_simulations}
We model the superfluid Fermi gas using a Density Functional Theory approach in the formulation known as Superfluid Local Density Approximation (SLDA)~\cite{Bulgac2012}. The energy of the system is computed as an integral over the energy density functional
\begin{equation}
    E(t)=\int
  \mathcal{E}[\rho(\bm{r},t),\tau(\bm{r},t),\bm{j}(\bm{r},t), \nu(\bm{r},t)]\,\mathrm{d}\bm{r}\label{eqn:E}
\end{equation}
which depends on the local densities: total ($\rho$), kinetic ($\tau$), current ($\bm{j}$) and anomalous ($\nu$), respectively. They are expressed via Bogoliubov amplitudes $\{u_n(\bm{r},t), v_n(\bm{r},t)\}$, where explicit forms for the most important ones are (see Ref.~\cite{Boulet2022} for others):
\begin{subequations}
  \label{eq:densities2}
  \begin{align}
    \rho
    &= 2\!\!\!\!\!\!\sum_{0<E_n<E_c}\!\!\left[
      \abs{v_{n}}^2 f(-E_n)+\abs{u_{n}}^2 f(E_n)
      \right], 
    \\
    \tau
    &= 2\!\!\!\!\!\!\sum_{0<E_n<E_c}\!\!\left[
      \abs{\bm{\nabla}v_{n}}^2 f(-E_n)+\abs{\bm{\nabla}u_{n}}^2 f(E_n)
      \right], \label{eq:n}
    \\
    \bm{j}
    &= \frac{2}{m}\!\!\!\!\!\!\sum_{0<E_n<E_c}\!\!\!\!\textrm{Im}\bigl[
         (v_{n}\bm{\nabla} v_{n}^*) f(-E_n)
         -  (u_{n}\bm{\nabla} u_{n}^*) f(E_n)
      \bigr]. \label{eq:j}
  \end{align}
\end{subequations}
Position and time dependence of functions have been omitted to shorten notation, the factor of 2 in these expressions accounts for spin degeneracy, $f(E) = 1/[e^{E/T}+1]$ is the Fermi-Dirac distribution function, and we work in units where $\hbar=k_B=1$.
We assume that the thermal occupation probabilities $f(E_n)$ do not change significantly in time, and we keep them as frozen over the entire evolution, set to the values defined by the initial conditions. This assumption is reasonable since the vortex dynamics is slow compared to the Fermi time scale $\eF^{-1}$, and the system always stays close to equilibrium. The densities are computed from all quantum states, with quasiparticle energies up to the cut-off energy $E_c\approx 10\eF$, and thus they account for both superfluid and normal components.
Note also that SLDA does not correspond to the commonly used Local Density Approximation (LDA), as the kinetic contribution $\tau$ is computed directly via derivative operators acting on quasiparticle wave functions.

The equations of motion are formally equivalent to time-dependent Bogoliubov-de Gennes (BdG) equations
\begin{gather}
  i\frac{\partial}{\partial t}
  \begin{pmatrix}
    u_n(\bm{r}, t)\\
    v_n(\bm{r}, t)
  \end{pmatrix}
  =
  \begin{pmatrix}
    \hat{h}(\bm{r}, t) & \Delta(\bm{r}, t)\\
    \Delta^*(\bm{r}, t) & -\hat{h}^*(\bm{r}, t)
  \end{pmatrix}
  \begin{pmatrix}
    u_n(\bm{r}, t)\\
    v_n(\bm{r}, t)
  \end{pmatrix},\label{eq:TDSLDA}
\end{gather}
with the difference that the single-particle Hamiltonian $\hat{h}$ and the pairing field $\Delta$ are related to energy density functional (\ref{eqn:E}) via appropriate functional derivatives:
\begin{equation}
    \hat{h} =-\frac{\nabla^2}{2m} + U(\bm{r}) +  V_{\textrm{ext}}(\bm{r}),\quad \Delta =-\frac{\delta\mathcal{E}}{\delta \nu^{*}}.
    \label{eq:single-part-ham}
\end{equation}
The mean-field $U=\delta\mathcal{E}/\delta \rho$ and pairing $\Delta$ potentials contain beyond mean-field corrections which depend on the interaction strength through the parameter $k_Fa$.
Finally, $U$ and $\Delta$ depend on the Bogoliubov amplitudes (via the respective densities), and the resulting equations of motion are highly nonlinear. 
Since in the computation we consider only states up to the cut-off energy $E_c$, the theory must be appropriately renormalized - see Ref.~\cite{Boulet2022} for the explicit expressions for the functional, the associated functional derivatives and renormalization scheme (in the computation we used a variant with the effective mass being equal to the bare mass.) 

The gas is confined by an external axially-symmetric potential $V_{\textrm{ext}}(r=\sqrt{x^2+y^2})$ which is zero for $r<R$ and rises smoothly but rapidly to a value much larger than the chemical potential for $r>R$. The radius of  the cylindrical potential is $R=90\,k_F^{-1}$. The lack of the potential dependence along the $z$-direction simplifies the computation since the Bogoliubov amplitudes factorize as $\{u_n(x,y), v_n(x,y)\}e^{ik_z z}$ with plane-waves along the $z$-direction such that the system reduces to a collection of 2D problems. We solve them on a spatial lattice of size $192\times 192$ (with lattice spacing $dx=k_F^{-1}$) for $16$ values of $k_z$. 
To initialize the time evolution, we use the phase imprinting method to find a numerical solution of the static variant of Eq.~(\ref{eq:TDSLDA}) (replacing $i\frac{\partial}{\partial t}$ with the eigenenergy $E_n$) which represents a quantum vortex located at position $(r_0,0)$.
In the computation, we use the W-SLDA Toolkit~\cite{WSLDAToolkit}, which is an advanced software for solving BdG-type problems. The code combines state-of-the-art algorithms (spectral methods for computation of derivatives, multistep Adams-Bashforth-Moulton of $5^{\textrm{th}}$ order integration scheme, error suppression techniques) with high-performance computing techniques~\cite{Wlazlowski:10.1093}. 
The significant computational advancements of the last decade in developing this high-precision toolkit led us to the level where the successful detection of the tiny cyclotron motion became possible.

\subsection{Vortex tracking}
\label{sub:Vortex_tracking}

We analyze the order parameter $\Delta(x,y)=\abs{\Delta(x,y)}e^{i\varphi(x,y)}$ to extract the position of the quantum vortex. Namely, the vortex core is localized at point $(x_\textrm{v},y_\textrm{v})$ around which the phase $\varphi$ rotates by $2\pi$ and $\abs{\Delta(x_\textrm{v},y_\textrm{v})}=0$. By employing the simplex algorithm of Nelder and Mead, implemented within GNU Scientific Library, we search for the zero of the function $\abs{\Delta(x,y)}$. To evaluate the order parameter for arbitrary coordinates ($x$,$y$), we use 
\begin{equation}
    \Delta(x,y) = \sum_{k_x, k_y}\Delta(k_x,k_y) e^{i(k_x x + k_y y)},
\end{equation}
where $\Delta(k_x,k_y)$ are discrete Fourier coefficients, obtained from the lattice data.  
For completeness, the order parameter is related to the Bogoliubov amplitudes by
\begin{equation}
    \Delta
    = -\frac{C}{\rho^{1/3}} \phantom{2}\!\!\!\!\!\!\sum_{0<E_n<E_c}\!\!\!
      u_{n}v_{n}^{*}
      \bigl[f(-E_n)-f(E_n)\bigr],
\end{equation}
where $C$ is a (negative) constant that depends on the interaction parameter $k_F a $ and the cut-off energy $E_c$~\cite{Boulet2022}.
The dependence of the pairing gap of the uniform system on $k_Fa$ provided by our DFT framework is consistent with ab-initio calculations and experimental measurements~\cite{Boulet2022}. In the BCS regime the bulk gap grows exponentially as $\bar{\Delta}_{\rm BCS}\approx \left(8/e^2\right) \eF \exp[\pi/(2\kF a)]$, while at unitarity it reaches $\bar{\Delta}_{\rm UFG}\approx 0.46\,\eF$ (see Ref.~\cite{Boulet2022} for the explicit interpolating formula).

The quality of data provided by the W-SLDA Toolkit and the tracking method allows us to identify the vortex position with sublattice resolution with uncertainty at the level of $1\%$ of the lattice spacing. We refer the reader to Sec.~\ref{sec:Ensuring} for further tests on the quality of the vortex trajectory extraction.

\begin{figure}[t]
    \centering \includegraphics[width=0.99\columnwidth]{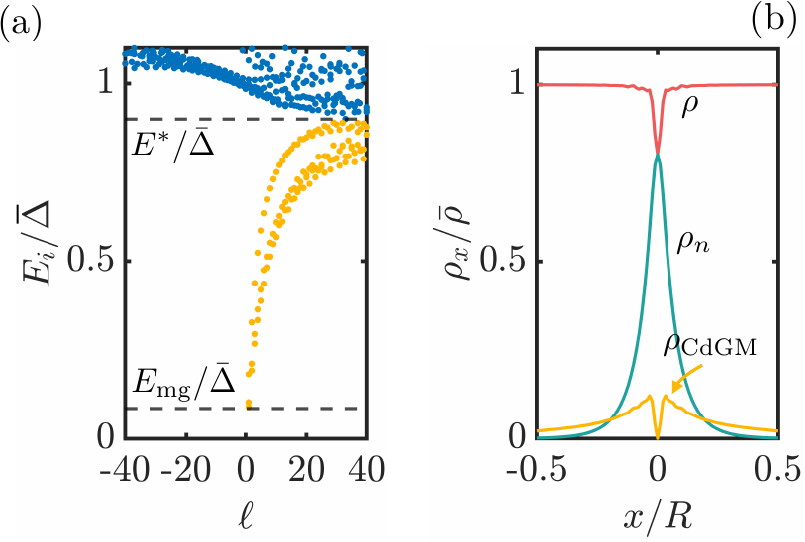}
    \caption{(a) Spectrum of the quasi-particle states as a function of the quantum number $\ell$. Yellow (blue) color corresponds to states whose energy is smaller (greater) than $E^*=0.9\,\bar{\Delta}$. (b) Cross-section of the density profiles along a path passing through the center, where $x=0$ denotes the vortex core.
    }
    \label{fig:Fig_CdG}
\end{figure}

\subsection{Normal component and CdGM states}
Remarkably, DFT simulations give also access to the properties of each quasiparticle state. Due to axial symmetry, the quasiparticle states for a centered vortex should have a well-defined angular momentum quantum number $\ell=\langle v_n|\hat{l}_z|v_n\rangle/\langle v_n|v_n\rangle$, where $v_n$ is the probability amplitude that the $n$-th quantum state is occupied by a particle. The energy spectrum shown in Fig.~\ref{fig:Fig_CdG}(a) features a bundle of states (depicted in yellow) within the energy gap $\bar{\Delta}$ (to be more precise, in the energy interval $0<E_i<E^*:=0.9\,\bar{\Delta}$). The numerically computed $\ell$'s for these states are integer numbers (up to $10^{-6}$ accuracy). Notice, in passing, that the energy of the first state, $E_1$ corresponds to the mini-gap energy $E_{\mathrm{mg}}=\bar{\Delta}^2/\varepsilon_F$ and that no states with $\ell=0$ exist in the gap, in agreement with previous studies~\cite{Magierski2022,Magierski2024}. These are the Caroli-de Gennes-Matricon (CdGM) vortex core states~\cite{Caroli1964}, which are distinguished from the continuum of states (depicted in blue) with energies $E_i>E^*$.  

It is worth stressing that $\rho_n$ is significantly different from the density of the CdGM core states, which at zero temperature is given by
\begin{equation}
 \rho_{\mathrm{CdGM}}(\bm{r})=2\sum_{E_n<E^*}|v_n(\bm{r})|^2.
\end{equation}
This appears clear in Fig.~\ref{fig:Fig_CdG}(b).  

\begin{figure*}[t]
    \centering
    \includegraphics[width=0.9\linewidth,trim={0 90 0 110},clip]{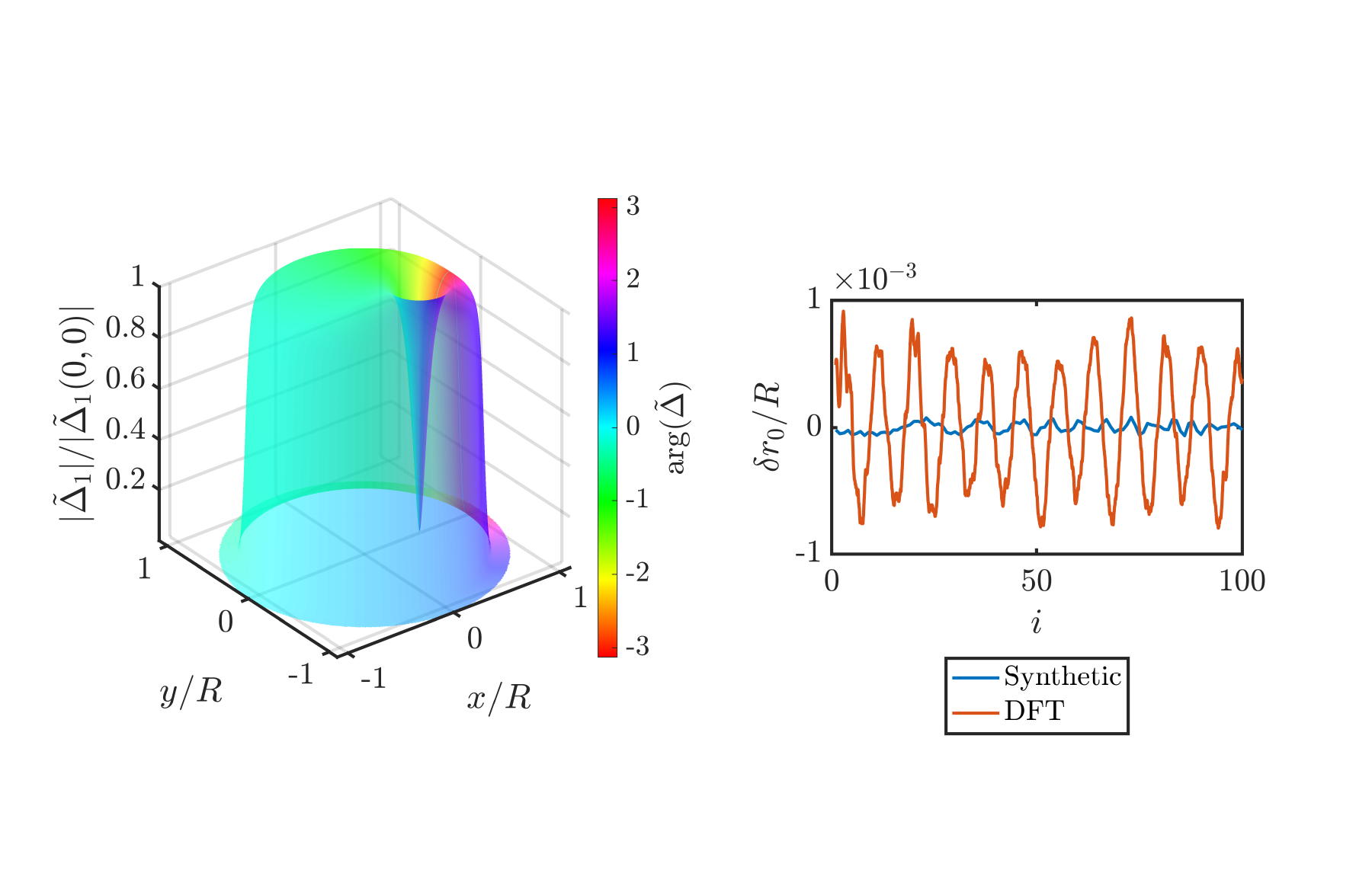}
    \caption{Left panel: plot of the synthetic state~(\ref{eq:Artificial_Delta}), at time $i=1$, used to benchmark the vortex-tracking algorithm employed to extract the vortex position $\bm{r}_0(t)$ from DFT simulations. Right panel: the position of the vortex extracted from our DFT simulations (red line) is characterized by regular oscillations significantly ($\sim 12$ times) wider than the irregular fluctuations (blue line) resulting from vortex tracking in synthetic states. For comparison, the radius of the box is $R=90\,dx$. }
    \label{fig:Fig_Supp_Mat}
\end{figure*}
\section{Ensuring the Robustness of Small-Amplitude Vortex Oscillations}
\label{sec:Ensuring}
The main technical challenge that we encountered during the development of our study was ensuring that the observed transverse oscillations $\delta r_0(t)$ of the vortex center (see Fig.~1.c and Fig.~3.b of the main text) were genuine signals rather than artifacts due to numerical noise. Recall, in fact, that these oscillations are definitely small, their typical amplitude being of the order of $10^{-3}\,R$, where $R = 90\,dx$ is the radius of the circular box and $dx$ is the mesh-element length (equivalently, the oscillations have an amplitude $\sim 0.1\,dx$). To prove that what we observed is not a numerical artifact we performed several tests on all the numerical tools involved in the extraction of $\delta r_0(t)$.

\subsection{Resolution of the vortex-tracking algorithm}
As explained in the Methods section of the main text, our vortex-tracking algorithm involves a suitable analysis of the superfluid order parameter $\Delta(x,y)$ based on the simplex algorithm of Nelder and Mead, implemented within GNU Scientific Library~\cite{Gough2009-vk}, combined with the spectral method for the interpolations. To evaluate the resolution of this tracking algorithm, we applied it to an artificial set of ``input states" $\{\tilde{\Delta}_i(x,y)\}$, where $i=1,\,\dots,\,100$ corresponds to different time samples. The structure
\begin{equation}
\begin{split}
    \tilde{\Delta}_i(x,y) &=\frac{1+\tanh\frac{R-\sqrt{x^2+y^2}}{\xi}}{2}\\
    &\times\left[1-\exp\left(-\frac{\sqrt{(x-x_0(i))^2+(y-y_0(i))^2}}{\xi} \right)\right]\\
    &\times\exp\left[i\left(\arctan\frac{y-y_0(i)}{x-x_0(i)} -\arctan\frac{y-y_0^\prime(i)}{x-x_0^\prime(i)}\right)\right]
    \label{eq:Artificial_Delta}
\end{split}
\end{equation}
of these synthetic states is illustrated in the left panel of Fig.~\ref{fig:Fig_Supp_Mat} and reproduces the main properties of the actual states $\{\Delta_i(x,y)\}$ generated by our DFT simulations. These include a conic-like dependence of $\tilde{\Delta}$ on $x$ and $y$ in a disk-like region of radius $\sim \xi$ centered at $(x_0,\,y_0)$, a phase defect therein, and a superfluid velocity tangent to the circular boundary of radius $R$ at any time $i$, hence the phase factor $\arctan[(y-y_0^\prime(i))/(x-x_0^\prime(i))]$, where  $x_0^\prime=x_0 R^2/(x_0^2+y_0^2)$ and $y_0^\prime=y_0 R^2/(x_0^2+y_0^2)$ . The position 
\begin{equation}
    x_0(i)=r_0\cos(\Omega i), \qquad y_0(i)=r_0\sin(\Omega i)
\end{equation}
of the physical vortex was evolved in such a way to imprint a smooth uniform circular motion of frequency $\Omega$. As a result, the set $\{\tilde{\Delta}_i(x,y)\}$ constitutes an established reference standard, being characterized by an ideally constant value of $r_0(i)=\sqrt{x_0(i)^2+y_0(i)^2}$.

\begin{figure}[b]
    \centering
    \includegraphics[width=0.99\linewidth]{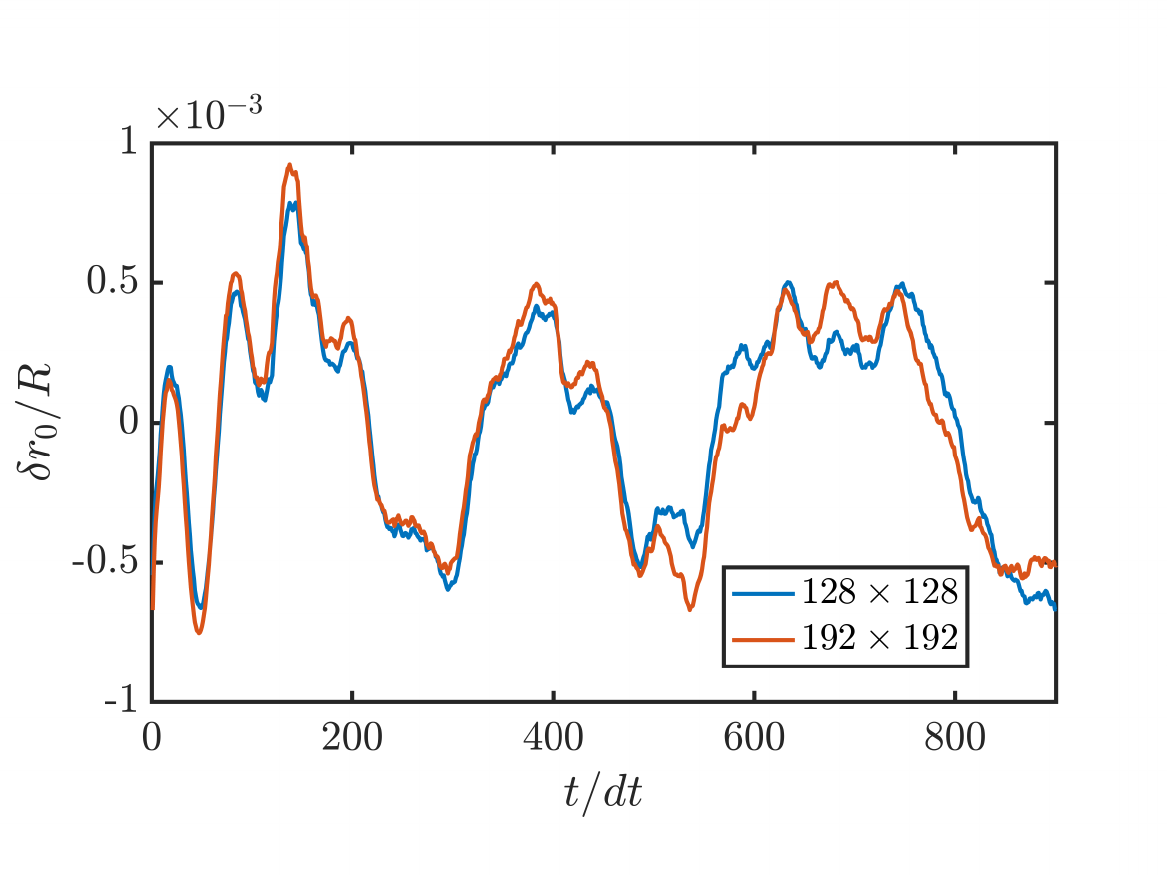}
    \caption{Time evolution of the vortex radial position for $k_F a= - 0.75$ and for two different spatial grids: 
    $128\times 128$ and  $192\times 192$ with lattice spacings $\kF^{-1}$ and $\frac{2}{3}\kF^{-1}$ respectively.
    The coarse-graining of the spatial grid minimally affects the properties of the vortex trajectory. }
    \label{fig:Compare_spatial_mesh}
\end{figure}
We sampled the so-defined set ${\tilde{\Delta}_i(x,y)}$ on a regular spatial grid of the same size as in the case of results from the main paper and used it as input for our vortex-tracking algorithm. The output, shown in the right panel of Fig.~\ref{fig:Fig_Supp_Mat} (blue line), is compared to a typical example of vortex oscillations (red line) from our DFT simulations. It is evident that the actual vortex oscillations are significantly ($\sim 12$ times) larger than the fluctuations associated with the synthetic states. The latter, in fact, should be regarded as a mere numerical noise that limits the resolution of our vortex-tracking algorithm. Additionally, the actual vortex oscillations visible in DFT simulations exhibit a clear sinusoidal behavior, unlike the irregular fluctuations associated with synthetic states, which contain multiple Fourier components. This benchmark contributes to proving that the reported vortex oscillations, though smaller than the spatial mesh element, are well above the resolution limit of our vortex-tracking algorithm.

\subsection{Amplifying and suppressing transverse oscillations}
To test that the observed cyclotron motion is due to the vortex mass, we repeat the time-dependent simulations with modified initial velocities for the vortex. 
The sensitivity of the vortex trajectory with respect to the initial velocity is a clear indicator that the equation of motion must be of the second order, as in Eq~(2).

\begin{figure}[t]
    \centering
    \includegraphics[width=0.99\columnwidth]{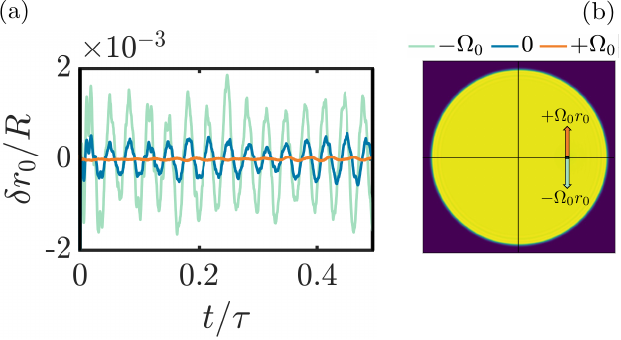}
    \caption{ 
    (a) Radial oscillations for three different initial velocities: $+\Omega_0r_0\hat{\phi}$ (orange line), zero (dark green),  and $-\Omega_0r_0\hat{\phi}$ (light green). (b) Illustration of the considered initial velocities for the vortex (the background color represents the total density $\rho(\bm{r})$ at $k_Fa=-0.75$). }
    \label{fig:Fig_Amplify_Suppress}
\end{figure}

This demonstration is achieved by generating the initial states in a rotating reference frame, where the original (single particle) Hamiltonian $\hat{h}$, defined in Eq.~(8), is replaced by $\hat{h}^\prime = \hat{h}-\Omega \hat{l}_z$ ($\hat{l}_z$ is the third component of the angular momentum operator). The parameter $\Omega$ represents the rotation frequency of the new frame and can be adjusted to control the vortex initial velocity in the original frame, given by $\dot{\bm{r}}_0 = \Omega r_0 \, \hat{\phi}$, ($\hat{\phi}$ being the azimuthal unit vector). This velocity determines the amplitude, but not the frequency, of the radial oscillations (since they are small, the anharmonic contributions do not manifest). According to the massive point-vortex model~\cite{Richaud2021}, there exists a specific angular velocity 
\begin{equation}
    \Omega_0 (r_0) = \frac{1}{\tau} \frac{2/(1-\tilde{r}_0^2)}{1+\sqrt{1-2\mathfrak{m}/(1-\tilde{r}_0^2) }}
    \label{eq:Omega_0}
\end{equation}
and thus an initial velocity $\dot{\bm{r}}_0$, for which the massive vortex exhibits a smooth, uniform circular motion. Any deviation from this condition triggers radial oscillations, with the amplitude increasing as the deviation becomes larger. The cases considered in the main text (Fig.~1) correspond to  $\Omega=0$, representing therefore a moderate perturbation with respect to $\Omega_0$. As illustrated in Fig.~\ref{fig:Fig_Amplify_Suppress} these oscillations  (dark green line) can be suppressed (orange line) upon setting $\Omega=+\Omega_0$ or amplified (light green line) upon setting $\Omega=-\Omega_0$. It clearly shows that different initial conditions result in oscillations with different amplitudes but sharing the same frequency. This is consistent with the predictions of the massive point vortex model~\cite{Richaud2021} and constitutes an important benchmark of our analysis.
Eventually, it is worth observing that the orange line of Fig.~\ref{fig:Fig_Amplify_Suppress}.a is characterized by residual fluctuations qualitatively and quantitatively similar to the ones associated with the synthetic states from Fig.~\ref{fig:Fig_Supp_Mat}. The amplitude of these residual fluctuations is about $0.01dx$. They can be, therefore, advisedly interpreted as numerical noise. Up to this precision, we are able to obtain the circular motion of the vortex if its initial velocity is appropriately chosen, which is perfectly consistent with the constant-value prediction of our massive point-vortex model~\cite{Richaud2021}.

\subsection{Impact of the spatial mesh and time step} 
We repeated our numerical simulations using different spatial meshes and time steps to consolidate and cross-check our results. More specifically, we demonstrated that the properties of vortex transverse oscillations are robust, not influenced by the choice of grid size, and not affected by temporal discretization errors. We tested grid sizes up to 1.5 times coarser and time steps up to 2 times larger than the default ones used to produce all our results. In both cases, decreasing the integration time step or refining the lattice spacing, as compared to the ones used in the main runs, do not induce significant changes. In Fig.~\ref{fig:Compare_spatial_mesh} we compare the time evolution of the vortex radial position for two meshes with fixed physical size but different lattice spacings.
The deviations between trajectories are smaller by at least one order of magnitude than the amplitude of the transverse oscillations. 
This strongly indicates that the physical phenomena at play are accurately captured, thus validating the reliability and accuracy of our simulations.

\subsection{Tracking vortex dipoles} 
\begin{figure}[h!]
    \centering
    \includegraphics[width=0.99\linewidth]{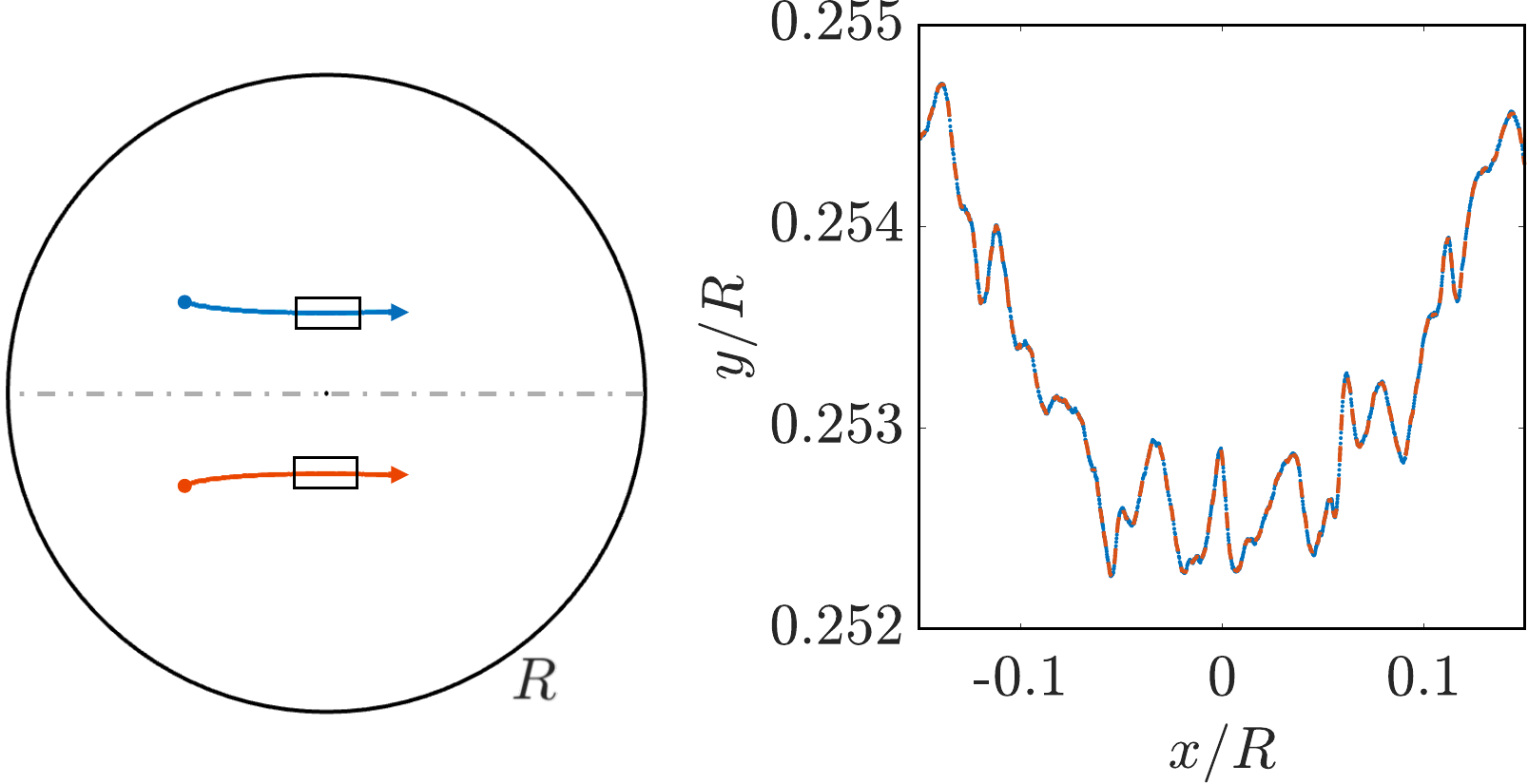}
    \caption{Left panel: Trajectories of a vortex-antivortex pair confined in a disk-like trap of radius $R$. Boxes denote the portions of trajectories that are zoomed and compared.  Right panel: comparison between the upper and the mirrored version of the lower box. The transverse oscillations of the two vortices are nearly perfectly symmetric with respect to the horizontal axis, a circumstance that further rules out numerical artifacts.}
    \label{fig:Vortex_dipole}
\end{figure}

To further validate the output of our vortex-tracking algorithm, we devised an additional test involving two vortices with opposite charges symmetrically placed with respect to the horizontal axis of the disk-like trap. According to predictions of our massive point-vortex model~\cite{Richaud2021}, the vortices should exhibit equal and opposite transverse oscillations. Time-dependent DFT simulations confirmed that their trajectories are nearly perfectly symmetric, with transverse oscillations in anti-phase (see Fig.~\ref{fig:Vortex_dipole}). To make the test even stronger, we rotated the horizontal axis (with respect to which we expect the symmetric evolution) by some angle. In this way, we have broken reflection symmetry related to the computational mesh. 
The strong (anti-)correlation between the two vortex-dynamics further rules out the possibility of numerical artifacts affecting the vortex transverse oscillations.

\end{document}